\documentclass{PoS}

\newcommand{\hess}{\textsc{H.E.S.S.}}
\newcommand{\hesstwo}{\textsc{HESS-II}}

\newcommand{\fermi}{{\textsl{Fermi}}/LAT}
\newcommand{\gray} {$\gamma$\mbox{-}ray} 
\newcommand{\grays}{$\gamma$\mbox{-}rays}
\newcommand{\esse} {$\rm 1ES \; 0414+009$}
\newcommand{\shbl} {$\rm SHBL \; J001355.9-185406$}
\newcommand{\pks}  {$\rm PKS \; 0447-439$}
\newcommand{\aplib}{$\rm AP \; Lib$}
\newcommand{\rxs}  {$\rm 1RXS \; J101015.9-311909$}
\newcommand{\es}   {$\rm 1ES \; 1312-423$}
\newcommand{\swiftxrt} {{\textsl{Swift}/XRT}}
\newcommand{\cena} {$\rm Centaurus \; A$}
\newcommand{\beppo} {{\textsl{Beppo}}SAX}

\title{New AGNs discovered by \hess}

\ShortTitle{New AGNs discovered by \hess}

\author{\speaker{Y. Becherini}$^{a,b}$ 
  B. Behera$^{c}$, J. Biteau$^{b}$, M. Cerruti$^{d}$, B. Giebels$^{b}$, 
  J.-P.~Lenain$^{d}$, M.~de~Naurois$^{b}$, 
  M.~Punch$^{a}$, M. Raue$^{e}$, D. Sanchez$^{f}$, F. Volpe$^{f}$, A. Zech$^{d}$
  on behalf of the H.E.S.S. Collaboration \\
  $^{a}$Astroparticule et Cosmologie (APC), CNRS, Universit\'e Paris Diderot -- Paris 7, Paris, France \\
  $^{b}$Laboratoire Leprince-Ringuet, Ecole Polytechnique, CNRS/IN2P3, Palaiseau, France \\   
  $^{c}$Landessternwarte, Universitat Heidelberg, Heidelberg, Germany \\
  $^{d}$Laboratoire Univers et Theories (LUTH), Observatoire de Paris, CNRS, Universit\'e Paris Diderot -- Paris 7, 
  Meudon, France \\
  $^{e}$Universitat Hambourg, Institut fur Experimentalphysik, Hambourg, Germany \\  
  $^{f}$Max-Planck-Institut fur Kernphysik, Heidelberg, Germany \\
  E-mail: \email{Yvonne.Becherini@apc.univ-paris7.fr}
}

\abstract{During the last year, six new Active Galactic Nuclei (AGN) have been discovered and studied by \hess\ 
  at Very High Energies (VHE). Some of these recent discoveries have been made thanks 
  to new enhanced analysis methods and are presented at this conference for the first time.  
  The three blazars \esse, \shbl\ and \rxs\
  have been targeted for observation due to their high levels of radio and X-ray fluxes,
  while the \fermi\ catalogue of bright sources triggered the observation of \pks\ 
  and $\rm AP \; Librae$. Additionally, the BL Lac \es\ was discovered in the field-of-view (FoV) of \cena\ 
  thanks to the large exposure dedicated by \hess\ to this particularly interesting source.
  The newly-discovered sources are presented here and in three companion presentations at this conference.
}

\FullConference{25th Texas Symposium on Relativistic Astrophysics - TEXAS 2010\\
		December 06-10, 2010\\
		Heidelberg, Germany} 

\begin{document}

\section{Introduction}
\label{Intro}

Eight years of operation of the High Energy Stereoscopic System (\hess) have been 
at the heart of a breakthrough in the field of \gray\ VHE astronomy: 
more than $70$ \gray\ sources have been discovered, 
$30$ of which are extra-galactic.
\hess\ will continue operation while awaiting the completion of the ongoing upgrade of the array, which will lead 
to the phase II of the experiment (see Sec. \ref{Conclu} for more details).
Meanwhile, \hess\ is improving its sensitivity also through the development and successful application 
of new enhanced analysis methods, such as for instance \cite{DeNaurois} and \cite{Becherini}.
These new analysis methods allow the detection of new faint \gray\ sources, having 
very low flux levels, i.e. of the order of $1\%$ C.U.\footnote{``Crab Units'', 
i.e., compared to a hypothetical source with the Crab nebula intensity and spectrum at the 
zenith angle of observation.} or less, in a moderate amount of observation time.
Six newly-discovered AGNs will be presented here, the detections of two of 
which (\rxs\ and \es) are announced at this conference for the first time.

\hess\ observation campaigns on AGNs have been largely based in the past 
on the criterion proposed in \cite{Costamante02}, where, briefly, 
a source is considered a good VHE candidate if it 
exhibits already high levels of both X-ray and radio emissions. 
More recently, a new complementary observation strategy is in place since the advent of the \fermi\ telescope, 
in which AGNs are targeted if the extrapolation of the GeV spectra (e.g. as given in \cite{Fermi})
to higher energies -- taking into account the \textsl{EBL\footnote{Extra-Galactic Background Light, resulting from
the combined emission of stars and galaxies from the earliest times of their formation to the present.}}
absorption as a function of redshift -- indicate that they may have a detectable counterpart in the VHE domain.
The spectra of distant blazars are modified by absorption of the VHE \grays\ on the EBL, 
with an effect which depends on the distance to the source and the EBL level assumed. 
Correspondingly, for a source at a known distance, if the instrinsic spectrum is assumed 
(e.g. by extrapolation from lower energies), then a limit can be placed on the EBL level, 
by evaluating the absorption required to produce the measured spectrum.

The two observation strategies above led to the recent discovery of 
several new AGNs of the BL Lacertae\footnote{
BL Lac objects are generally characterized by relativistic jets beamed towards the observer, 
which are thought to originate the large and rapid variability seen in all energy ranges. 
Additionally, these objects show high polarization and lack of emission lines.} (BL Lac or blazar) type. 
The three blazars \esse, \shbl\ and \rxs\ 
were targeted for observations following \cite{Costamante02},
while the \fermi\ catalogue criterion motivated the observation of \pks\ 
and \aplib. 
Moreover, low-redshift X-ray blazars were also proposed to be good candidates for VHE astronomy by \cite{Stecker}, 
and hence some such were also included in the list of possible interesting targets.
\es\ is one such AGN, whose location is about $2^\circ$ from the position of the radio-galaxy 
\cena, a source which has been extensively observed by \hess\ and discovered in VHE \grays. 

Table 1 summarizes the results presented at this conference in this and the companion proceedings 
(see references in the table). The main characteristics of the AGNs presented
(position, redshift, flux level, mean observing angle, number of standard deviations of the detection, observing strategy) 
are given, together with the corresponding good quality livetime, which represents the amount of good quality data 
after weather selection criteria and corrected for the instrument dead-time. 

\section{\esse: one of the most distant VHE blazars with a known redshift}

\esse\ is one of the distant sources having a known redshift of about $\sim 0.29$, 
which has been detected by \hess\
with a significance of about $7.8 \; \sigma$
after a four-year observing campaign, leading to a total good livetime of about $74$ hours 
(see \cite{ES0414} at this conference).  
The detection of this faint source ($0.5\%$ C.U.) occurred at the same time
the \fermi\ found its GeV emission by analyzing the 
$21$ months dataset; this led to a Astronomer's Telegram \cite{ATel0414} in which the
discovery of the GeV-TeV emissions was announced simultaneously.
This AGN is a high-frequency-peaked BL Lac (HBL), where the
high-energy part of the spectral energy distribution (SED) shows a peak 
of the emission above a few TeV. 
Such a feature is difficult to explain with a standard one-zone SSC scenario, 
therefore, this source belongs to the class of the hard TeV BL Lac objects, which are challenging for modelling. 

\begin{table}[t]
\begin{center} 
\begin{small}
\begin{tabular}{|l|c|c|c|c|c|c|c|c|}
\hline
\hline
Source name & RA & Dec & $z$ & LT   & $\phi_0$ & $\theta_{\rm zen}$ & N$_{\sigma}$ & Strat. \\
\hline
\hline
\esse\ \bf{\cite{ES0414}}  & 64.22  & 1.08   & $0.287$      & $74$ & $0.5\%$  & $26^\circ$   & $7.8$ & \cite{Costamante02} \\
\shbl\                 & 3.48   & -18.90 & $0.095$      & $38$ & $\sim1\%$  & $13^\circ$   & $\sim5$ & \cite{Costamante02} \\
\pks\  \bf{\cite{PKS0447}} & 72.35  & -43.84 & $[0.176,0.5]$  & $14$ & $4.5\%$  & $23^\circ$   & $14$ & \cite{Fermi} \\
\aplib\ \bf{\cite{APLIB}}  & 229.42 & -24.37 & $0.049$ & $11$ & $2\%$  & $13^\circ$   & $7$ & \cite{Fermi} \\
\rxs\                 & 152.57 & -31.32 & $0.14$       & $33$ & $\sim2.5\%$ & $13^\circ$   & $7.2$ & \cite{Costamante02}\\
\es\                  & 198.76 & -42.61 & $0.105$      & $168^{*}$ & $0.4\%$ & $24^\circ$   & $6.8$ & \cite{Stecker} \\
\hline
\hline
\end{tabular}
\label{TablePerformance}
\caption{\small{The six new AGNs detected by \hess\ during the past year. 
    The columns present the source name and corresponding references, 
    position in right ascension and declination in degrees, 
    the redshift $z$, observation livetime in hours, 
    flux in C.U., mean observing angle in degrees, 
    significance in number of standard deviations, 
    and finally the references used to guide the observing strategies.
    The sources detailed in other proceedings at this conference, 
    have this reference in bold besides their names.   
    (*) For \es\ this is the total livetime, but correcting for the camera acceptance this reduces to about $65$ hours.
}}
\end{small}
\end{center} 
\end{table} 

\section{Discovery of \shbl: first \hess, then \fermi}

\begin{figure}[t]
\centerline{
  \includegraphics[width=6.7cm]{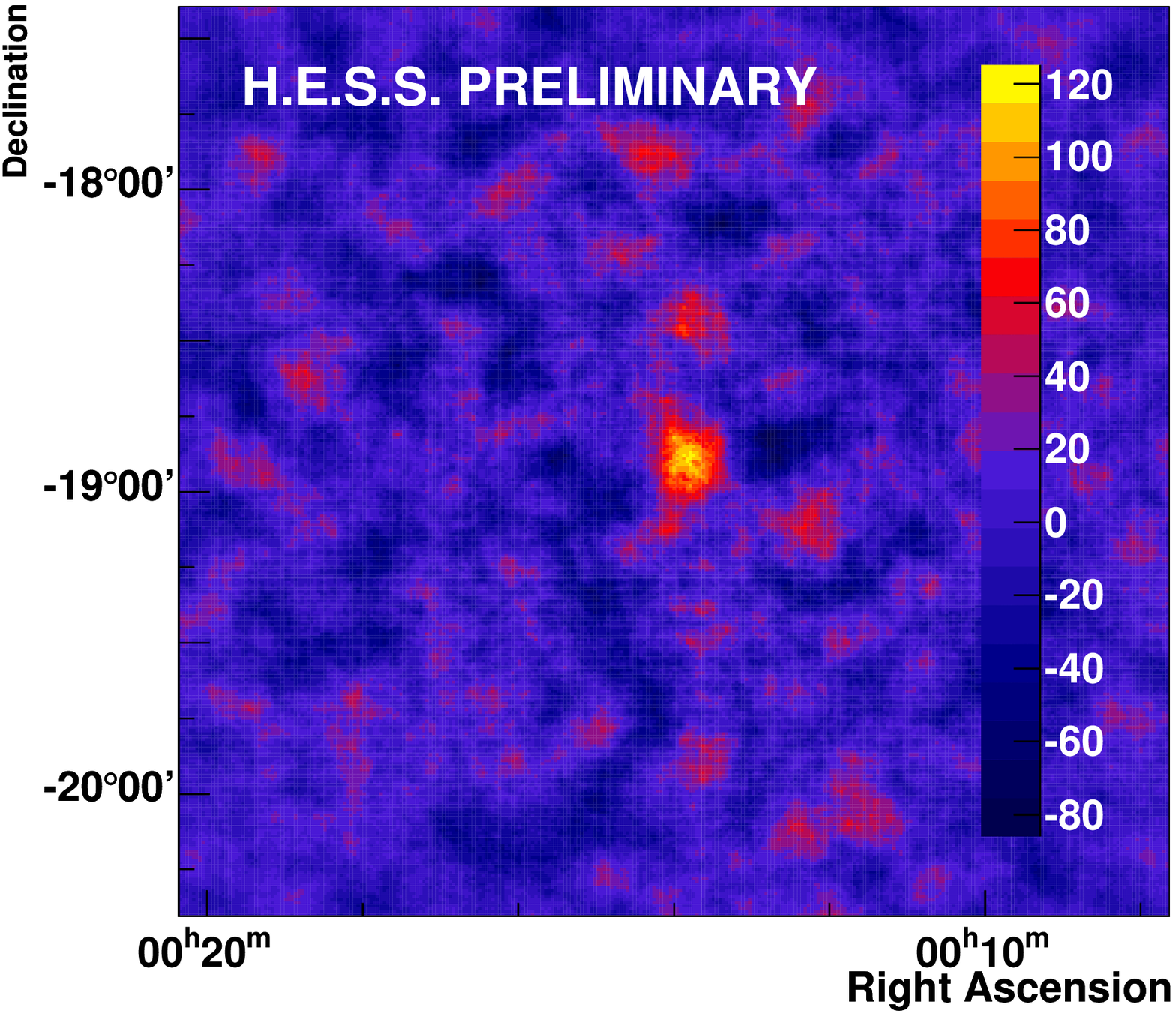}
  \includegraphics[width=6.7cm]{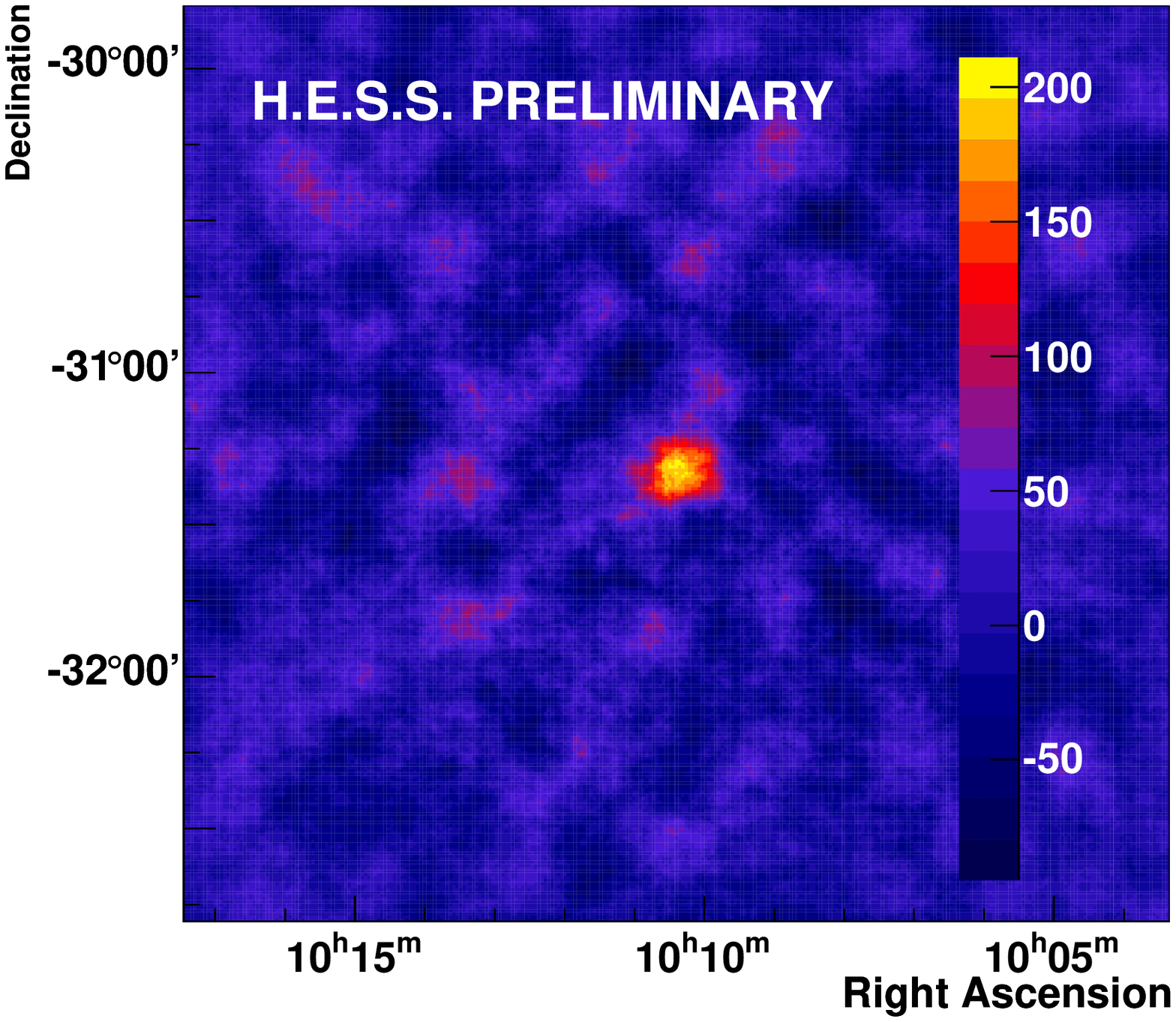}
}
  \caption{\small{{\textsl{Left panel.}} Excess map of the region including the blazar \shbl.
  {\textsl{Right panel.}} Excess map of the region including the blazar \rxs.
  }}
  \label{1355}
\end{figure}

The newly-discovered \shbl, also belonging to the HBL category, 
is much closer, with a redshift of 0.095.
The source is contained among the sources of the RASS/BSC\footnote{ROSAT All 
Sky Survey Bright Source catalogue ($18811$ sources)} 
of soft ($0.1-2 \rm \; keV$) X-ray sources (see \cite{Voges}), 
where its flux was found to be $1.01\times10^{-11} \rm erg/cm^{2}/s$ and is in the 
NVSS catalogue\footnote{NRAO (National Radio Astronomy Observatory) VLA (Very Large Array) 
Sky Survey ($1.8 \times 10^6$ sources)} of radio sources at 1.4 GHz (see \cite{Condon}) 
which gives its flux density at $\rm 29.5 \; mJy$.
The source was selected according to the criterion in \cite{Costamante02}, but also 
from the list of $150$ HBL sources in the SHBL\footnote{Sedentary High-frequency BL Lacertae} 
catalogue (see \cite{Giommi}), where, briefly, interesting VHE targets show an extremely high ratio 
of X-ray to radio fluxes.
The source was not present in the first year \fermi\ catalogue, 
indicating that its flux in the GeV range was below the 1-year sensitivity.
\hess\ discovered the VHE emission from this BL Lac object \cite{ATel1355} 
using the two advanced analysis methods mentioned in Sec.\ \ref{Intro}, 
with a statistical significance of about $5 \; \sigma$ above $\rm 300 \; GeV$ 
in 38 hours of good quality observations carried out between July 2008 and August 2010, 
see Fig.~\ref{1355}a for the excess map. 
The light curve of the source is found to be compatible with a constant flux at the level of $\sim 1\%$ C.U.
The source was monitored in optical wavelengths 
over the same period by ATOM\footnote{Automatic Telescope for Optical Monitoring}, 
operated by the \hess\ collaboration and located next to the \hess\ site. 
Additionally, \swiftxrt\ \cite{Swift} observations in the $\rm 0.2 - 10 \; keV$ energy 
range were triggered by \hess\ in September 2010. 
No significant variability in either optical or X-ray bands down to a timescale of a few days has been found.
Subsequent to the \hess\ discovery, the \fermi\ Collaboration announced the detection of a source positionally 
consistent with the BL Lac with a significance of $7 \; \sigma$, see \cite{Sanchez}. 
The \fermi\ preliminary analysis 
indicated a faint hard-spectrum source with a 2-year averaged flux above $100 \; \rm MeV$ 
of $(0.9 \pm 0.7) \times 10^9$ $\rm ph/cm^{2}/s$ 
and a photon index of $1.5 \pm 0.2$, considering statistical errors only.

\section{\pks: trying to constrain the redshift}

\pks\ is one of the brightest hard-spectrum blazars in the \fermi\ catalogue. 
\hess\ observed this source for a total livetime of about $14$ hours, announcing the 
discovery of the VHE \gray\ emission in \cite{ATel0447} with a significance of $14 \; \sigma$. 
The redshift of this source is not well known; a value of $0.205$ was initially proposed by \cite{RedshiftPerlman}, 
whereas a more recent study \cite{RedshiftLandt} provides only a lower limit of $0.176$. 
The study detailed in \cite{PKS0447} at this conference 
reverses the procedure for the evaluation of the EBL level discussed in Sec.\ \ref{Intro}.
It supposes the EBL to be now known with some certitude and explores the range of acceptable
redshifts for which the extrapolation of the spectrum from the \fermi\ range does not overproduce 
the predicted spectrum in the \hess\ range, within measurement errors. 
In this way, an upper limit in redshift of $0.5$ was derived 
at the $95\%$ confidence level.

\section{$\rm AP \; Librae$: first LBL of the southern hemisphere}

AP Librae (\aplib) is the closest of the new blazars presented here,
at a redshift of $0.049$, and was targeted for \hess\ observations 
thanks to the \fermi\ bright source catalogue. 
This AGN has been discovered by \hess\ after $11$ hours of livetime 
with a significance of $7 \; \sigma$ and announced in \cite{ATel1514}.
The SED (see \cite{APLIB}, at this conference) 
of this source shows that \aplib\ is a low-frequency peaked BL Lac (LBL) and  
presents an extremely broad high-energy component which extends from X-ray to VHE,
showing that, for this source, synchrotron X-ray brightness is not required to produce VHE \grays\ and that  
an external Compton component may be responsible for detected VHE emission.
\aplib\ is the third LBL detected at VHE and the first one discovered in the southern hemisphere.

\section{Discovery of VHE \grays\ from the BL Lac object \rxs}

\rxs\ is a blazar at a redshift of $0.14$. 
The source is a HBL contained in the RASS/BSC catalogue
of soft X-ray sources, where its flux is given as 
$2.9 \times 10^{-11} \; \rm erg/cm^{2}/s$ and is in the NVSS catalogue of radio sources at 
$\rm 1.4 \; GHz$, where its flux density is 74.3 mJy.
This AGN has been shown to be a good candidate for VHE Cherenkov astronomy also for its 
extremely high X-ray to radio flux ratio, pointed out in the SHBL catalogue, and 
was not present in the first year \fermi\ catalogue (\cite{Fermi}).
\hess\ observed this AGN in a dedicated campaign between 2007 and 2010 at
a mean observing angle of $13^\circ$, 
leading to a total good livetime of $33$ hours. 
The analysis of this AGN was carried out with the two independent enhanced analysis procedures mentioned 
in Sec.\ \ref{Intro}, which give consistent results.
VHE emission from the BL Lac object \rxs\ has thus been discovered by \hess\ 
with an excess of $233$ \gray\ events, at a significance level of $7.2 \; \sigma$,
and is presented at this conference for the first time. 
The VHE \gray\ excess sky map is shown in Fig.\ \ref{1355}b.
The flux of the source is about $2.5\%$ C.U., and spectral studies are ongoing.

\begin{figure}[t]
\centerline{
  \includegraphics[width=6.5cm]{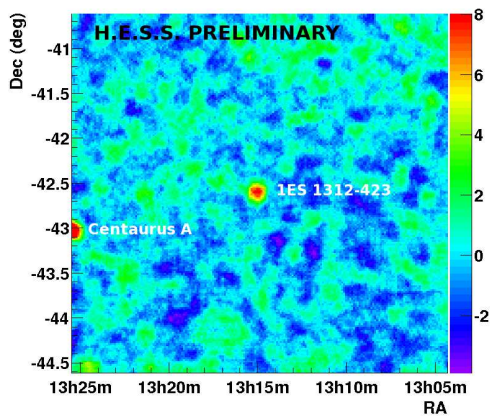}
}
  \caption{\small{Significance map of \es. The source on the left-hand side is the radio-galaxy \cena.}}
  \label{1312}
\end{figure}

\section{Discovery of \es: in the same FoV of \cena}

\es\ is a blazar at a redshift of 0.105 present in the Einstein slew survey sample \cite{Perlman}, 
in the RASS/BSC, and in the \beppo\ spectral survey of BL Lacs \cite{Beckmann}. 
The source was proposed as a good VHE $\gamma$-ray candidate by \cite{Stecker}, 
where it was suggested through a simple SSC model that low-redshift X-ray BL Lac objects
should have a VHE counterpart. The source was not present in the \fermi\ first year catalogue.
\es\ lies at the edge of the FoV of \hess\ for observations centred around the radio-galaxy \cena.
\hess\ carried out a deep observation of \cena\, leading to its discovery in the VHE domain
in early 2009 (\cite{CenA}), reporting a flux level of $0.8\%$ in C.U. 
For this reason, the data set available for the entire FoV is large, and has been useful 
to perform a dedicated analysis on this blazar, despite the lower collection efficiency 
at about $2^\circ$ distance from the observing position.
The source has been analyzed with the two above-mentioned enhanced analysis methods leading to the
the discovery of its VHE emission with a $6.8$ $\sigma$ significance 
in $168$ hours of good livetime taken between 2004 and 2010. 
The total corrected livetime, taking into account that the observations were carried out 
at large offset with respect to the source position, is of about $65$ hours. 
Preliminary studies of the light curve show no detectable variability, 
similarly to observations in other wavelengths, which were not variable either.

\section{Conclusions}

\label{Conclu}

\hess\ is now in the process of refurbishing the four mirror surfaces and building 
a new very-large telescope, in preparation for 
the operation of its second phase, which will begin in 2012. 
The new telescope will permit to lower the threshold to about $30 \; \rm GeV$ in the single-telescope mode, 
opening a complementary observational window to high-energy phenomena.
As negligible \gray\ absorption is expected in the $30-100 \; \rm GeV$ 
energy range, therefore more distant objects will become visible, enriching the extra-galactic 
source catalogue and permitting to put stronger constraints on the 
EBL. By enriching the array with the addition of a fifth telescope at the centre, 
the experiment will be about a factor of two more sensitive with respect
to the present configuration in multi-telescope mode.
In the meantime, \hess\ has succeeded in enhancing its sensitivity uniquely by the development 
and successful application of new, more powerful analysis methods. 
Thanks to the enhanced sensitivity provided by these new methods, 
sources which were not seen with the standard analyses,
became detectable.  
At this conference, six newly-detected AGNs have been presented, the discovery of two of which (\rxs\ and \es) 
has been shown for the first time.
While the detections of \pks\ and \aplib\ were motivated by the \fermi\ catalogue of bright sources, 
the discovery of \shbl\ was first achieved at VHE by \hess, and subsequently by \fermi, 
integrating over a larger dataset.  
The SEDs of \esse\ and \aplib\ have peculiar features which are 
difficult to explain with standard one-zone SSC scenarios,
giving intriguing inputs for the modelling of LBLs and HBLs.
With the detection of \esse\ and \es, two blazars having flux levels of $0.5\%$ and $0.4\%$ C.U., 
\hess\ enriches the catalogue of the weakest-ever sources detected in VHE \grays. 
These recent discoveries demonstrate that the field of extragalactic VHE \gray\ astronomy is very active 
and will become more so with the advent of the second phase of \hess: \hesstwo.

\section{Acknowledgements}

The support of the Namibian authorities and of the University of Namibia in facilitating 
the construction and operation of \hess\ is gratefully acknowledged, 
as is the support by the German Ministry for Education and Research (BMBF), 
the Max Planck Society, the French Ministry for Research, 
the CNRS-IN2P3 and the Astroparticle Interdisciplinary Programme of the CNRS, 
the U.K. Science and Technology Facilities Council (STFC), the IPNP of the Charles University, 
the Polish Ministry of Science and Higher Education, 
the South African Department of Science and Technology and National Research Foundation, 
and by the University of Namibia. We appreciate the excellent work of the technical support staff in Berlin, 
Durham, Hamburg, Heidelberg, Palaiseau, Paris, Saclay, and in Namibia in the construction and operation of the equipment.

\end{document}